
\input harvmac

\def\journal#1&#2(#3){\unskip, \sl #1\ \bf #2 \rm(19#3) }
\def\andjournal#1&#2(#3){\sl #1~\bf #2 \rm (19#3) }

\def\ie{{\it i.e.}}
\def\eg{{\it e.g.}}

\def\frac#1#2{{#1\over#2}}

\def\half{\frac12}

\def\inbar{\,\vrule height1.5ex width.4pt depth0pt}
\def\IC{\relax\hbox{$\inbar\kern-.3em{\rm C}$}}
\def\IR{\relax{\rm I\kern-.18em R}}
\def\IP{\relax{\rm I\kern-.18em P}}
\def\IZ{\relax{\rm I\kern-.18em Z}}

%
%

%
\catcode`\@=11
\def\slash#1{\mathord{\mathpalette\c@ncel{#1}}}
\overfullrule=0pt

\def\CC{{\cal C}}

\def\underrel#1\over#2{\mathrel{\mathop{\kern\z@#1}\limits_{#2}}}

\catcode`\@=12


%

\def\tr{{\rm tr}}


\def\tn{\widetilde{N}}
\def\tw{\widetilde{W}}
\def\tr{\widetilde{R}}
\def\ts{\widetilde{S}}

\Title{ \rightline{}} {\vbox{\centerline{Fundamental Strings and
Black Holes}}}
\medskip
\centerline{\it Amit Giveon${}^{1}$ and David Kutasov${}^{2}$}
\bigskip
\smallskip
\centerline{${}^{1}$Racah Institute of Physics, The Hebrew
University} \centerline{Jerusalem 91904, Israel}
\smallskip
\centerline{${}^2$EFI and Department of Physics, University of
Chicago} \centerline{5640 S. Ellis Av., Chicago, IL 60637, USA }

\bigskip\bigskip\bigskip
\noindent

We propose a black hole thermodynamic description of highly excited
charged and uncharged perturbative string states in $3+1$
dimensional type II and $4+1$ dimensional heterotic string theory.
We also discuss the generalization to extremal and non-extremal
black holes carrying magnetic charges.

\vfill \Date{11/06}

\lref\WaldYP{
R.~M.~Wald,
``Quantum field theory in curved space-time and black hole thermodynamics,''
University of Chicago Press (1994).
}

\lref\MaldacenaKY{
J.~M.~Maldacena,
``Black holes in string theory,''
arXiv:hep-th/9607235.
}

\lref\GrossAR{
D.~J.~Gross and P.~F.~Mende,
``String Theory Beyond the Planck Scale,''
Nucl.\ Phys.\ B {\bf 303}, 407 (1988).
}

\lref\MendeWT{
P.~F.~Mende and H.~Ooguri,
``Borel Summation of String Theory for Planck Scale Scattering,''
Nucl.\ Phys.\ B {\bf 339}, 641 (1990).
}

\lref\PeetHN{
A.~W.~Peet,
``TASI lectures on black holes in string theory,''
arXiv:hep-th/0008241.
   }

\lref\DasSU{
S.~R.~Das and S.~D.~Mathur,
``The quantum physics of black holes: Results from string theory,''
Ann.\ Rev.\ Nucl.\ Part.\ Sci.\  {\bf 50}, 153 (2000)
[arXiv:gr-qc/0105063].
}

\lref\AharonyTI{
O.~Aharony, S.~S.~Gubser, J.~M.~Maldacena, H.~Ooguri and Y.~Oz,
``Large N field theories, string theory and gravity,''
Phys.\ Rept.\  {\bf 323}, 183 (2000)
[arXiv:hep-th/9905111].
}

\lref\McClainID{
B.~McClain and B.~D.~B.~Roth,
``Modular Invariance for Interacting Bosonic Strings at Finite Temperature,''
Commun.\ Math.\ Phys.\  {\bf 111}, 539 (1987).
}

\lref\OBrienPN{
K.~H.~O'Brien and C.~I.~Tan,
``Modular Invariance Of Thermopartition Function And Global Phase Structure
Of Heterotic String,''
Phys.\ Rev.\ D {\bf 36}, 1184 (1987).
}

\lref\AtickSI{ J.~J.~Atick and E.~Witten, ``The Hagedorn Transition
and the Number of Degrees of Freedom of String Theory,'' Nucl.\
Phys.\ B {\bf 310}, 291 (1988).
}

\lref\PolchinskiZF{
J.~Polchinski,
``Evaluation Of The One Loop String Path Integral,''
Commun.\ Math.\ Phys.\  {\bf 104}, 37 (1986).
}

\lref\DabholkarYR{
A.~Dabholkar,
``Exact counting of black hole microstates,''
Phys.\ Rev.\ Lett.\  {\bf 94}, 241301 (2005)
[arXiv:hep-th/0409148].
}

\lref\DabholkarDQ{
A.~Dabholkar, R.~Kallosh and A.~Maloney,
``A stringy cloak for a classical singularity,''
JHEP {\bf 0412}, 059 (2004)
[arXiv:hep-th/0410076].
}

\lref\KrausVZ{
P.~Kraus and F.~Larsen,
``Microscopic black hole entropy in theories with higher derivatives,''
JHEP {\bf 0509}, 034 (2005)
[arXiv:hep-th/0506176].
}

\lref\SenIZ{
A.~Sen,
``Entropy function for heterotic black holes,''
JHEP {\bf 0603}, 008 (2006)
[arXiv:hep-th/0508042].
}

\lref\KrausZM{
P.~Kraus and F.~Larsen,
``Holographic gravitational anomalies,''
JHEP {\bf 0601}, 022 (2006)
[arXiv:hep-th/0508218].
}

\lref\ElitzurCB{
S.~Elitzur, A.~Forge and E.~Rabinovici,
``Some global aspects of string compactifications,''
Nucl.\ Phys.\ B {\bf 359}, 581 (1991).
}

\lref\MandalTZ{
G.~Mandal, A.~M.~Sengupta and S.~R.~Wadia,
Mod.\ Phys.\ Lett.\ A {\bf 6}, 1685 (1991).
}

\lref\WittenYR{
E.~Witten,
``On string theory and black holes,''
Phys.\ Rev.\ D {\bf 44}, 314 (1991).
}

\lref\McGuiganQP{
M.~D.~McGuigan, C.~R.~Nappi and S.~A.~Yost,
``Charged black holes in two-dimensional string theory,''
Nucl.\ Phys.\ B {\bf 375}, 421 (1992)
[arXiv:hep-th/9111038].
}

\lref\HorneGN{
J.~H.~Horne and G.~T.~Horowitz,
``Exact black string solutions in three-dimensions,''
Nucl.\ Phys.\ B {\bf 368}, 444 (1992)
[arXiv:hep-th/9108001].
}

\lref\GiveonMI{
A.~Giveon, D.~Kutasov, E.~Rabinovici and A.~Sever,
``Phases of quantum gravity in AdS(3) and linear dilaton backgrounds,''
Nucl.\ Phys.\ B {\bf 719}, 3 (2005)
[arXiv:hep-th/0503121].
}

\lref\KutasovCT{
D.~Kutasov,
``A geometric interpretation of the open string tachyon,''
arXiv:hep-th/0408073.
}

\lref\GiveonZZ{
A.~Giveon and A.~Sever,
``Strings in a 2-d extremal black hole,''
JHEP {\bf 0502}, 065 (2005)
[arXiv:hep-th/0412294].
}

\lref\VenezianoZF{
G.~Veneziano,
``A Stringy Nature Needs Just Two Constants,''
Europhys.\ Lett.\  {\bf 2}, 199 (1986);
``String-theoretic unitary S-matrix at the threshold of black-hole
production,''
JHEP {\bf 0411}, 001 (2004)
[arXiv:hep-th/0410166].
}

\lref\SusskindWS{
L.~Susskind,
``Some speculations about black hole entropy in string theory,''
arXiv:hep-th/9309145.
}

\lref\HorowitzNW{
G.~T.~Horowitz and J.~Polchinski,
``A correspondence principle for black holes and strings,''
Phys.\ Rev.\ D {\bf 55}, 6189 (1997)
[arXiv:hep-th/9612146];
``Self gravitating fundamental strings,''
Phys.\ Rev.\ D {\bf 57}, 2557 (1998)
[arXiv:hep-th/9707170].
}

\lref\DamourAW{
T.~Damour and G.~Veneziano,
``Self-gravitating fundamental strings and black holes,''
Nucl.\ Phys.\ B {\bf 568}, 93 (2000)
[arXiv:hep-th/9907030].
}

\lref\MohauptJD{
T.~Mohaupt,
``Strings, higher curvature corrections, and black holes,''
arXiv:hep-th/0512048.
}

\lref\KarczmarekBW{
J.~L.~Karczmarek, J.~M.~Maldacena and A.~Strominger,
``Black hole non-formation in the matrix model,''
JHEP {\bf 0601}, 039 (2006)
[arXiv:hep-th/0411174].
}

\lref\HorneCN{
J.~H.~Horne, G.~T.~Horowitz and A.~R.~Steif,
``An Equivalence between momentum and charge in string theory,''
Phys.\ Rev.\ Lett.\  {\bf 68}, 568 (1992)
[arXiv:hep-th/9110065].
}

\lref\DabholkarJT{
A.~Dabholkar and J.~A.~Harvey,
``Non-renormalization of the Superstring Tension,''
Phys.\ Rev.\ Lett.\  {\bf 63}, 478 (1989).
}

\lref\DabholkarYF{
A.~Dabholkar, G.~W.~Gibbons, J.~A.~Harvey and F.~Ruiz Ruiz,
``Superstrings and Solitons,''
Nucl.\ Phys.\ B {\bf 340}, 33 (1990).
}

\lref\MaldacenaCG{
J.~M.~Maldacena and A.~Strominger,
``Semiclassical decay of near-extremal fivebranes,''
JHEP {\bf 9712}, 008 (1997)
[arXiv:hep-th/9710014].
}

\lref\KutasovZH{
D.~Kutasov, F.~Larsen and R.~G.~Leigh,
``String theory in magnetic monopole backgrounds,''
Nucl.\ Phys.\ B {\bf 550}, 183 (1999)
[arXiv:hep-th/9812027].
}

\lref\BehrndtEQ{
K.~Behrndt, G.~Lopes Cardoso, B.~de Wit, D.~Lust, T.~Mohaupt and W.~A.~Sabra,
``Higher-order black-hole solutions in N = 2 supergravity and Calabi-Yau
string backgrounds,''
Phys.\ Lett.\ B {\bf 429}, 289 (1998)
[arXiv:hep-th/9801081].
}

\lref\LopesCardosoWT{
G.~Lopes Cardoso, B.~de Wit and T.~Mohaupt,
``Corrections to macroscopic supersymmetric black-hole entropy,''
Phys.\ Lett.\ B {\bf 451}, 309 (1999)
[arXiv:hep-th/9812082].
}

\lref\LopesCardosoCV{
G.~Lopes Cardoso, B.~de Wit and T.~Mohaupt,
``Deviations from the area law for supersymmetric black holes,''
Fortsch.\ Phys.\  {\bf 48}, 49 (2000)
[arXiv:hep-th/9904005].
}

\lref\LopesCardosoUR{
G.~Lopes Cardoso, B.~de Wit and T.~Mohaupt,
``Macroscopic entropy formulae and non-holomorphic corrections for
supersymmetric black holes,''
Nucl.\ Phys.\ B {\bf 567}, 87 (2000)
[arXiv:hep-th/9906094].
}

\lref\LopesCardosoXN{
G.~Lopes Cardoso, B.~de Wit and T.~Mohaupt,
``Area law corrections from state counting and supergravity,''
Class.\ Quant.\ Grav.\  {\bf 17}, 1007 (2000)
[arXiv:hep-th/9910179].
}

\lref\MohauptMJ{
T.~Mohaupt,
``Black hole entropy, special geometry and strings,''
Fortsch.\ Phys.\  {\bf 49}, 3 (2001)
[arXiv:hep-th/0007195].
}

\lref\LopesCardosoQM{
G.~Lopes Cardoso, B.~de Wit, J.~Kappeli and T.~Mohaupt,
``Stationary BPS solutions in N = 2 supergravity with R**2 interactions,''
JHEP {\bf 0012}, 019 (2000)
[arXiv:hep-th/0009234].
}

\lref\LopesCardosoFP{
G.~Lopes Cardoso, B.~de Wit, J.~Kappeli and T.~Mohaupt,
``Examples of stationary BPS solutions in N = 2 supergravity theories  with
R**2-interactions,''
Fortsch.\ Phys.\  {\bf 49}, 557 (2001)
[arXiv:hep-th/0012232].
}

\lref\KutasovRR{
D.~Kutasov,
``Accelerating branes and the string / black hole transition,''
arXiv:hep-th/0509170.
}

\lref\GiveonPX{
A.~Giveon and D.~Kutasov,
``Little string theory in a double scaling limit,''
JHEP {\bf 9910}, 034 (1999)
[arXiv:hep-th/9909110].
}

\lref\FZZ{V.~A.~Fateev, A.~B.~Zamolodchikov and Al.~B.~Zamolodchikov,
unpublished.}

\lref\KutasovPF{
D.~Kutasov,
``Irreversibility Of The Renormalization Group Flow In Two-Dimensional
Quantum Gravity,''
Mod.\ Phys.\ Lett.\ A {\bf 7}, 2943 (1992)
[arXiv:hep-th/9207064].
}

\lref\HsuCM{
E.~Hsu and D.~Kutasov,
``The Gravitational Sine-Gordon model,''
Nucl.\ Phys.\ B {\bf 396}, 693 (1993)
[arXiv:hep-th/9212023].
}

\lref\SahooPM{
B.~Sahoo and A.~Sen,
``alpha' corrections to extremal dyonic black holes in heterotic string
theory,''
arXiv:hep-th/0608182.
}

\lref\BalasubramanianEE{
V.~Balasubramanian and F.~Larsen,
``Near horizon geometry and black holes in four dimensions,''
Nucl.\ Phys.\ B {\bf 528}, 229 (1998)
[arXiv:hep-th/9802198].
}

\lref\KazakovPM{
V.~Kazakov, I.~K.~Kostov and D.~Kutasov,
``A matrix model for the two-dimensional black hole,''
Nucl.\ Phys.\ B {\bf 622}, 141 (2002)
[arXiv:hep-th/0101011].
}

\lref\GiveonJV{
A.~Giveon and D.~Kutasov,
``The charged black hole / string transition,''
JHEP {\bf 0601}, 120 (2006)
[arXiv:hep-th/0510211].
}

\lref\PolchinskiRQ{
J.~Polchinski,
``String theory. Vol. 1: An introduction to the bosonic string,''
``Vol. 2: Superstring theory and beyond,''
Cambridge University Press, 1998.}

\lref\CornalbaHC{
  L.~Cornalba, M.~S.~Costa, J.~Penedones and P.~Vieira,
  ``From fundamental strings to small black holes,''
  arXiv:hep-th/0607083.
}

\newsec{Introduction}

In quantum gravity in asymptotically flat $3+1$ dimensional
spacetime, generic high energy states with vanishing
charge\foot{There are generalizations to states with charge, angular
momentum, as well as to other dimensions of spacetime.} are believed
to be described by the Schwarzschild geometry. Their entropy is
expected to be given by the Bekenstein--Hawking (BH) formula
\eqn\bhen{S=A/4G_N~,}
with $A=4\pi R_h^2$ the area of the horizon, $R_h$ the Schwarzschild
radius and $G_N$ the Newton constant (see \eg\ \WaldYP\ for a
review).

This description is thermodynamic in nature. The Euclidean black
hole solution contributes to the canonical free energy at a
temperature equal to the Hawking temperature of the black hole. The
Minkowski solution can be thought of as an average over all states
with energy equal to the mass of the black hole. Thus, it
corresponds to the microcanonical ensemble. The entropy -- energy
relation implied by \bhen\ is $S=4\pi G_N M^2$. This is the leading
term in an asymptotic expansion in inverse energy. Corrections are
due to quantum and other effects.

The problem of providing a statistical interpretation to black hole
thermodynamics has received a lot of attention over the years and
much progress has been achieved (see \eg\
\refs{\WaldYP\MaldacenaKY\PeetHN-\DasSU} for reviews). In
particular, for gravity in asymptotically anti-de-Sitter spacetime,
which is dual to a field theory \AharonyTI, the entropy of large
$AdS$--Schwarzschild black holes is expected  (and in some examples
was verified) to agree with the high energy density of states of the
dual field theory. In other cases, such as large Schwarzschild black
holes in asymptotically flat spacetime, the nature of the
``microstates'' that lead to the BH entropy is not understood.

String theory contains gravity, so the above discussion applies to
it. When the string coupling $g_s$ is small, the theory has two
widely separated scales. One is $l_p$, at which quantum gravity
effects become important. The other is $l_s=\sqrt{\alpha'}=1/M_s$,
at which string corrections become important. For example, in $3+1$
dimensions one has $l_p=g_sl_s$, so $l_p\ll l_s$ at weak coupling.

In weakly coupled string theory, the Schwarzschild solution
describes the thermodynamics for energies $E\gg M_s/g_s^2$. In this
regime the Schwarzschild radius is large ($R_h\gg l_s$) and gravity
is reliable, but the spectrum of the theory is not understood. On
the other hand, for $El_s$ large but finite in the limit $g_s\to 0$
the spectrum is known, and one can study the corresponding
statistical mechanics using standard tools. To leading order in
$g_s$, the free energy is given by the string partition sum with
Euclidean time compactified on a circle of circumference
$\beta=1/T$, evaluated on a worldsheet torus
\refs{\PolchinskiZF\McClainID\OBrienPN-\AtickSI}. This approach is
statistical in nature, since the partition sum is obtained by
tracing over microstates.

A natural question is whether there is a description of the perturbative
string spectrum in terms of black hole thermodynamics. It
is reasonable to expect that such a description exists since for
$El_s\gg1$ the free string entropy is large and a thermodynamic
description should be appropriate. Indeed, for some classes of
perturbative heterotic string states with mass equal to charge,
thermodynamic descriptions in terms of extremal black holes were
proposed before
\refs{\DabholkarYR\DabholkarDQ\KrausVZ\SenIZ-\KrausZM}.

The main purpose of this note is to propose a thermodynamic
description for a class of perturbative string states with generic
mass and charges in $3+1$ dimensional type II and $4+1$ dimensional
heterotic string theory. According to this description, as one
approaches a highly excited fundamental string of mass $M$, the
angular sphere shrinks and decouples from the radial direction and
time. The latter are described by a two dimensional geometry with
asymptotically flat metric and linear dilaton in the radial direction.
The full near-horizon geometry is described by the (charged or uncharged)
two dimensional black hole
\refs{\ElitzurCB\MandalTZ\WittenYR\McGuiganQP\HorneGN-\GiveonMI}.
We show that the thermodynamic entropy of the black hole matches that
of the corresponding perturbative string states to leading order in
$M_s/M$ and any mass to charge ratio.

The above construction seems to be special to $3+1$ dimensions in
type II and $4+1$ dimensions in heterotic string theory. In other
dimensions one should also be able to replace highly excited strings
by geometries, but these geometries might be more complicated. An
important feature of our proposal is the decoupling of the angular
sphere from the radial coordinate and time in the near-horizon
geometry. This decoupling probably does not occur in general and the
near-horizon geometry is described by a highly curved background
which involves all $d$ dimensions of spacetime.

We also discuss the generalization of the above construction to the
case where in addition to the electric charges carried by
perturbative strings we turn on magnetic charges associated with
Neveu-Schwarz fivebranes and Kaluza-Klein monopoles. We construct
the exact worldsheet background corresponding to systems with
generic values of these four charges and mass, and study their
thermodynamics.

\newsec{Type II strings}

We start with type II string theory on
\eqn\geomthreeone{\IR^{3,1}\times \CC_6~,}
where $\CC_6$ is a compact manifold. The sigma model on $\CC_6$ is a
unitary $N=1$ superconformal field theory with central charge $c=9$
and a discrete spectrum of scaling dimensions starting at zero. We
are interested in the thermodynamics of highly excited perturbative
string states in this background. In this section we will provide
such a description, first for uncharged states and then for states
carrying up to two charges associated with a circle in $\CC_6$.

\subsec{Type II strings as black holes}

One way to approach the problem is to start with a large
Schwarzschild black hole, with Schwarzschild radius $R_h\gg l_s$,
and ask what happens when its mass decreases. For the Schwarzschild
solution, one has $R_h=2G_NM$, so as $M$ decreases $R_h$ does as
well. Eventually it becomes of order $l_s$ and string corrections
become important. Formally, this happens when $M\sim M_s/g_s^2$,
which is the transition region between the black hole and string
regimes \refs{\VenezianoZF\SusskindWS\HorowitzNW-\DamourAW}. Below
that mass, the size of the typical state is larger than the horizon
radius, but one can still hope to use a black hole picture to study
the thermodynamics.

As $G_N M/l_s\to 0$ we expect the BH temperature to approach the
Hagedorn one, and black hole thermodynamics to match smoothly to
perturbative string thermodynamics. In particular, for $M\gg M_s$
and $g_s\to 0$ the black hole entropy should behave like
\eqn\bbhh{S=\beta_H M~,}
with
\eqn\bbbhhh{\beta_H=2\sqrt{2}\pi l_s}
the inverse Hagedorn temperature of perturbative type II strings.

This behavior should be reproduced by the $\alpha'$ corrected
near-horizon Schwarzschild geometry. To find this geometry by
systematically including these corrections is a formidable task,
both because the perturbative corrections to the Einstein equations
are known only to the first few orders in $\alpha'$, and because
non-perturbative effects might be important.  We will next make a
proposal for its form.

For large values of the radial coordinate $r$, the space around a
small black hole is flat $(\IR^3)$. As $r$ decreases, the radius of
the angular two-sphere $S^2$ shrinks. We will {\it assume} that near
the horizon the geometry factorizes into a $1+1$ dimensional
background describing $(t,r)$ and a two-dimensional one for the
$S^2$. The main motivation for this assumption is that this is
expected to happen for certain charged extremal four dimensional
black holes that preserve some supersymmetry
(see e.g. \MohauptJD\ for a recent review). As we will see later,
it is then natural to expect that it happens for non-extremal and
uncharged black holes as well.

At any rate, under the above assumption the worldsheet CFT
corresponding to $S^2$ is trivial for the following reason. By
construction, this CFT must have an $SO(3)$ global symmetry. This
implies the existence of a worldsheet current $(J^a,\bar J^a)$, with
$a=1,2,3,$ in the adjoint of $SO(3)$, satisfying the conservation
equation
\eqn\jjaa{\bar\partial J^a+\partial\bar J^a=0~.}
Since the worldsheet theory on $S^2$ is a unitary CFT with a
discrete spectrum, one can show in general that $J^a$ and $\bar J^a$
must be separately conserved, \ie\ $\bar\partial J^a=\partial\bar
J^a=0$. If one of the currents is non-zero, the other should be
non-zero as well since the worldsheet CFT corresponding to the small
black hole background should be left-right symmetric. The symmetry
is thus enhanced to $SU(2)_L\times SU(2)_R$ with $J^a$ and $\bar
J^a$ satisfying the corresponding affine Lie algebra relations. As
is well known, backgrounds with such symmetries correspond
geometrically to $S^3$ rather than $S^2$, and contain fluxes of the
NS B-field through the sphere, which makes them inappropriate here.
The currents $J^a$ and $\bar J^a$ must thus vanish, so the simplest
possibility is that the worldsheet theory corresponding to $S^2$ is
trivial.

We conclude that the near-horizon geometry of the small black hole
contains only the radial direction and time. The corresponding
worldsheet CFT should have the same central charge as that
describing the flat spacetime at infinity. Since time translation
invariance is a symmetry, the simplest possibility is that the
near-horizon geometry contains a dilaton which is asymptotically
linear in the radial direction $\phi$, a function of $r$ that
corresponds near the boundary of this geometry to a canonically
normalized scalar field.

The slope of the dilaton, $Q$, determines the central charge of the
worldsheet theory for $\phi$ via the relation
\eqn\cphi{c_\phi=1+3Q^2~.}
Comparing the central charge of the theory of $(t,\phi)$ to that of
$\IR^{3,1}$ fixes $Q$. One finds that $Q=1$, such that the total
central charge of $(t,\phi)$ and their worldsheet superpartners is
equal to six.

One can think of the appearance of the linear dilaton as an analog
of gravitational RG flow. It is well known (see \eg\
\refs{\KutasovPF,\HsuCM}) that if one couples a two dimensional
theory which interpolates between UV and IR fixed points along an RG
flow to worldsheet gravity, the RG flow proceeds as a function of
the Liouville coordinate $\phi$. The boundary of space, $\phi\to\infty$,
where the string coupling goes to zero, corresponds to the UV fixed point,
while $\phi\to-\infty$ corresponds to the IR fixed point. In our case, the
analog of the Liouville coordinate is the radial coordinate $r$ or $\phi$.
On its own, the sigma model on the angular two-sphere in $\IR^3$
naturally goes to smaller central charge, and is massive in the infrared.
The radial direction compensates by developing a
non-trivial dilaton and increasing its central charge.

To summarize, we propose that the asymptotic form of the
near-horizon geometry of small black holes (\ie\ those with masses
in the perturbative string regime) is
\eqn\nearhor{\IR_t\times\IR_\phi\times\CC_6~.}
The dilaton is linear in the radial coordinate $\phi$ such that this
background is critical. All excitations in the near-horizon geometry
\nearhor\ are singlets of $SO(3)$ (\ie\ s-waves). A related fact is
that if the full background \geomthreeone\ preserves some
supersymmetry, only half of the generators are visible in the
near-horizon geometry \nearhor. The other half act trivially on all
degrees of freedom there, like the generators of $SO(3)$.

The small black hole must correspond to a two dimensional black hole
in $\IR_t\times\IR_\phi$. The unique solution to the string
equations of motion with the right properties is the
$SL(2,\IR)/U(1)$ black hole \refs{\ElitzurCB\MandalTZ-\WittenYR},
which is described by the metric and dilaton (here we set
$\alpha'=2$)
\eqn\sltwobh{\eqalign{ &ds^2=f^{-1}d\phi^2-fdt^2~, \qquad
f=1-{2M\over \rho}~,\cr &Qe^{-2\Phi}=Qe^{Q\phi}=\rho~.\cr }}
The dilaton is asymptotically linear in $\phi$, and approaches a
constant,
\eqn\dilhor{e^{-2\Phi_h}={2M\over Q}~,}
at the horizon. The level of $SL(2)$, $k$, is in general related to
the linear dilaton slope via the relation
\eqn\qqqq{Q=\sqrt{2\over k}~.}
Note that $k$ is the total level of $SL(2,\IR)$. The worldsheet
theory is superconformal; it consists of a bosonic $SL(2,\IR)$ WZW
model of level $k+2$, and three free fermions in the adjoint
representation that contribute $-2$ to the level. In our case $Q=1$,
so $k=2$. This value lies above the transition of
\refs{\KarczmarekBW,\GiveonMI}, so the black
hole is a normalizable state in the near-horizon geometry.

The Euclidean black hole is obtained by Wick rotating $t\to it$ in
\sltwobh. This gives rise to a cigar geometry, and one can read off
the inverse temperature of the black hole from the circumference of
Euclidean time at infinity. The result is
\eqn\betah{\beta_H=2\pi l_s\sqrt k~,}
which corresponds to a Hagedorn entropy at high energies $E\gg M_s$,
\eqn\hagent{S_{BH}=\beta_H E~.}
This result is valid for all $k$, despite the fact that \sltwobh\
was obtained by solving the equations of motion of dilaton gravity,
which are only valid for large $k$ (or small $Q$). This can be shown
by analyzing the exact spectrum of the coset theory $SL(2,\IR)/U(1)$
algebraically. From the sigma model point of view, it follows from
the fact that the coset preserves $N=2$ supersymmetry, which leads
to non-renormalization of the background \sltwobh.

Altogether, we conclude that the small black hole that provides a
thermodynamic description of uncharged fundamental string states
with mass $M\gg M_s$ in $\IR^{3,1}\times\CC_6$ is
\eqn\nearhorr{{SL(2,\IR)_2\over U(1)}\times\CC_6~.}
This black hole has a Hagedorn entropy \hagent. Its Hagedorn
temperature  \betah\ coincides\foot{The fact that the Little String
Theory entropy agrees with the perturbative string one for $k=2$, as
well as some other relations between the two problems, were pointed
out in \KutasovCT.} with that of perturbative fundamental strings
\bbbhhh.

The assertion that perturbative string states develop a linear
dilaton throat in their vicinity sounds at first sight surprising.
Note that this is only expected to occur for states with
sufficiently small mass. The black hole geometry \sltwobh, in which
the dilaton decreases as one moves away from the horizon, attaches
to flat space at the place where $e^\Phi$ reaches its asymptotic
value, $g_s$. In order for it to be valid in a significant range of
distances, the string coupling at the horizon, \dilhor, must be much
larger than the asymptotic coupling, or:
\eqn\rangemm{M\ll {M_s\over g_s^2}~.}
Thus, as one would expect, our description is only valid well below
the string/black hole correspondence region of
\refs{\VenezianoZF\SusskindWS\HorowitzNW-\DamourAW}.

{}From the point of view of the full geometry, the throat \sltwobh\
occupies a string size region around $r=0$. A simple way to see that
is to note that in \sltwobh\ the two-sphere has already disappeared.
Thus, \sltwobh\ must attach to the rest of the geometry in a region
where the size of the two-sphere is of the order of the string
scale. Another, heuristic, way to estimate the radial size of the
throat is to note that the geometry \sltwobh\ is reminiscent of the
Schwarzschild one, with the replacement $\rho\to r/g_s^2$. In the
coordinate $\rho$ the transition between the linear dilaton throat
and flat spacetime occurs at $\rho\sim 1/g_s^2$, which is equivalent
to $r\simeq 1$.

As the mass $M$ of the string state increases towards $M\sim M_s/g_s^2$,
the linear dilaton region shrinks and eventually disappears. On the other
hand, as $M/M_s$ decreases the throat formally becomes longer, but the
string coupling at the horizon grows and for $M\simeq M_s$ the theory
becomes strongly coupled. This is natural from the point of view of
string thermodynamics since the entropy of the corresponding
fundamental string states is small and one expects large
fluctuations in the thermal description.

One can also ask what happens for other dimensions of spacetime. The
general considerations above suggest that one should still be able
to replace fundamental strings with $M\gg M_s$ by a geometry with a
horizon. However, in this case the description of the near-horizon
region by an $SL(2,\IR)/U(1)$ black hole is inconsistent with the
free string entropy. We suspect that the origin of the problem is
the assumption of decoupling of the angular sphere $S^{d-2}$ from
the radial coordinate and time. In general, there is no motivation
for assuming this decoupling, and without it it is difficult to
determine the near-horizon geometry. The situation for states with
non-zero angular momentum in $3+1$ dimensions is also more
complicated due to the absence of $SO(3)$ symmetry in the
corresponding black hole solution.

\subsec{Charged strings as black holes}

In this subsection we will generalize the discussion of the previous
subsection to fundamental string states which carry momentum $n$ and
winding $w$ around a circle of radius $R$. Thus, we will take the
geometry \geomthreeone\ to have the form
\eqn\mins{\IR^{3,1}\times S^1\times \CC_5~.}
The left and right-moving momentum of the string on the $S^1$ is
given by
\eqn\qqnw{(q_L,q_R)=\left({n\over R}+{wR\over\alpha'}\, ,\, {n\over
R}-{wR\over\alpha'}\right)~.}
The mass-shell condition is
\eqn\mnnqq{\alpha' M^2=4N_L+\alpha' q_L^2=4N_R+\alpha' q_R^2~,}
where $N_L$, $N_R$ are the left and right-moving oscillator levels.

For large $N_L$ and/or  $N_R$, the entropy of free strings with mass
$M$ and charges $(q_L,q_R)$ is given by
\eqn\smqlqr{
S=2\pi\sqrt{2}\left(\sqrt{N_L}+\sqrt{N_R}\right) =\pi l_s\sqrt
2\left(\sqrt{M^2-q_L^2}+\sqrt{M^2-q_R^2}\right)~.}
For $q_L=q_R=0$ we saw in the previous subsection that the black
hole background \nearhorr\ provides a thermodynamic description of
these states. Black holes with generic $(q_L,q_R)$ are obtained by
adding a circle to the uncharged black hole, performing a boost
along it, followed by T-duality and another boost.
This leads \HorneCN\
to the background
\eqn\cbh{{SL(2,\IR)_2\times U(1)\over U(1)}\times\CC_5~.}
The charge to mass ratio of the black hole determines the way the
$U(1)$ in the denominator is embedded in $SL(2,\IR)_2\times U(1)$.
Denoting by $(J^3,\bar J^3)$ the left and right-moving currents in a
space-like Cartan subalgebra of  $SL(2,\IR)_L\times SL(2,\IR)_R$,
and by $(J,\bar J)$ the currents corresponding to the $U(1)$ factor
in the numerator of \cbh, the left and right components of the
gauged $U(1)$ current $(J_L,J_R)$ are given by
\eqn\jjjj{\eqalign{J_L=&J^3\cos\alpha_L +J\sin\alpha_L ~,\cr
J_R=&\bar J^3\cos\alpha_R  +\bar J\sin\alpha_R ~,\cr}} where
\eqn\alar{\eqalign{\sin\alpha_L=&{q_L\over M}~,\cr
\sin\alpha_R=&{q_R\over M}~.\cr }} In the special case $q_L=q_R=0$,
we have $\alpha_L=\alpha_R=0$, and the background \cbh\ reduces to
the uncharged two dimensional black hole \sltwobh, \nearhorr, with
$\CC_6=S^1\times \CC_5$.

For generic $(q_L,q_R)$, the ${SL(2,\IR)\times U(1)\over U(1)}$
factor in \cbh\ describes time, the radial coordinate of the four
dimensional charged black hole and the circle under which the
fundamental strings are charged. The angular two-sphere decouples in
the near-horizon region, as in the uncharged case.

The three dimensional background in \cbh\ is a special case of a
more general class of charged black holes of the form
\eqn\cbhk{{SL(2,\IR)_k\times U(1)\over U(1)}~.}
The geometry and thermodynamics of these black holes were studied in
\GiveonMI. Their entropy is
\eqn\sbh{S_{BH}=\pi l_s\sqrt
k\left(\sqrt{M^2-q_L^2}+\sqrt{M^2-q_R^2}\right)~.}
As in the uncharged case, one can obtain \sbh\ by analyzing the
background \cbhk\ algebraically. One can also describe this
background as a solution to dilaton gravity coupled to gauge fields
and use it to study the thermodynamics. Apriori this analysis is
only valid for large $k$, but in fact it is expected to be exact, as
for $q_L=q_R=0$.

For $k=2$, \sbh\ is equal to the free string entropy \smqlqr\ and we
propose that the background \cbh\ is in fact the near-horizon
geometry of the corresponding string states. Like \smqlqr, \sbh\ is
valid to leading order in $M_s/M$ and for arbitrary charge to mass
ratio $\alpha_L$, $\alpha_R$ \alar.

In general, the black hole \cbh\ is non-extremal and breaks all
supersymmetry. In some special cases, which correspond to extremal
black holes, part of the supersymmetry is preserved by the solution.
In particular, for
\eqn\mmqq{M=|q_R|}
and generic $q_L$ the solution preserves a quarter of the
supercharges of the background \nearhor\ and provides a
thermodynamic description of the corresponding $1/4$ BPS
Dabholkar-Harvey states \refs{\DabholkarJT,\DabholkarYF}. In this
case, $\alpha_L$ in \alar\ is generic, while $|\sin\alpha_R|=1$.
Thus, the right-moving component of \jjjj\ lies purely in the $S^1$
in \mins, \cbh, while $J_L$ acts both on the $SL(2,\IR)$ and the
$U(1)$ in \cbh.

For large $k$, one can think of \cbhk\ as a sigma model on
$AdS_2\times S^1$, whose properties can be obtained from the results
of appendix C of \GiveonMI. The radii of the AdS space and the
circle are given by
\eqn\rrads{R_{\rm AdS}={l_s\over 2}\sqrt{k}~,}
and
\eqn\rrs{R=l_s\sqrt{\left|{n\over w}\right|}~.}
The gauge fields associated with $G$ and $B$ on the circle have
constant field strengths on $AdS_2$ (proportional to $n$ and $w$).
The two dimensional dilaton takes the value
\eqn\diltwo{{1\over g_2^2}=\sqrt{k|nw|}~.}
In the case of interest to \cbh, $k=2$, the background \cbhk\ is
highly stringy, but as in the other cases discussed above one can
still formally continue the sigma model picture to this regime,
and the results \rrads, \rrs\ do not receive $\alpha'$
corrections. String loop corrections are small for $|nw|\gg1$, the
regime of interest.

A further restriction to the case
\eqn\mqlqr{M=q_L=-q_R~,}
corresponding to a string with winding but no momentum around $S^1$,
leads to a black hole that preserves half of the supersymmetry. From
\jjjj, \alar\ we see that in this case the gauging does not act on
the $SL(2,\IR)$ in \cbh, and the resulting background is
\eqn\nearext{SL(2,\IR)_2\times \CC_5~.}
The three dimensional string coupling can be obtained by combining
\rrs, \diltwo:
\eqn\gthree{{1\over g_3^2}=\sqrt{k}|w|=\sqrt{2}|w|~.}
Thus, string theory in the background \nearext\ is weakly coupled
for large $|w|$, and can be studied using standard perturbative
string theory techniques.

\subsec{States carrying electric and magnetic charges}

So far we focused on states which carry only electric charges \qqnw\
on the $S^1$ in \mins. In this subsection we generalize the
discussion to states that carry magnetic charges as well. To do
that, we take the compact manifold $\CC_5$ in \mins\ to have the
form $\CC_5=\ts^1\times\CC_4$, so that the geometry \mins\ is
\eqn\minss{\IR^{3,1}\times S^1\times \ts^1\times\CC_4~,}
and add to the strings with momentum and winding $(n,w)$ on $S^1$
discussed above $\tw$ $NS5$-branes wrapped around $S^1\times \CC_4$
and $\tn$ KK monopoles extended in the same five directions and
charged under the Kaluza-Klein gauge field associated with $\ts^1$.
The magnetically charged objects are BPS and we consider excitations
of this configuration with energy $E\gg M_s$, as before.

To compute the entropy of such states we need the corresponding
near-horizon geometry, which is given by
\eqn\cftii{{SL(2,\IR)_k\times U(1)\over U(1)}\times {SU(2)_k\over
Z(\tn)_L}\times \CC_4~.}
Here
\eqn\kkkii{k=\tn\tw~,}
and the $U(1)$ quotient acts as in \jjjj, \alar. As before, $k$
\kkkii\ is the total level of $SL(2,\IR)$ which receives
contributions of $k+2$ and $-2$ from bosons and fermions
respectively. Similarly, for the $SU(2)$ component in \cftii\ the
total level \kkkii\ receives a contribution of $k-2$ from a bosonic
$SU(2)$ WZW model and $+2$ from three free fermions in the adjoint
representation that are needed for superconformal symmetry.

To prove \cftii\ one can proceed as follows. For the special case
$q_L=q_R=0$, the background \cftii\ reduces to
\eqn\cftnew{{SL(2,\IR)_k\over U(1)}\times S^1\times {SU(2)_k\over
Z(\tn)_L}\times \CC_4~,}
which is known to describe near-extremal $NS5$-branes  and KK
monopoles  wrapped around $S^1\times\CC_4$
\refs{\MaldacenaCG,\KutasovZH}. The $Z(\tn)_L$ quotient in \cftnew\
is associated with the KK monopoles. It acts holomorphically on the
worldsheet; see \eg\ \KutasovZH\ for a more detailed discussion. The
value of the dilaton at the horizon is determined by the energy
density, as in \dilhor, \MaldacenaCG.

To generalize to the case of non-vanishing electric charges one can
perform the sequence of boosts and T-duality on the $S^1$ discussed
for the case of vanishing magnetic charges above. This leads to the
background \cftii\ whose entropy is given by \sbh, \kkkii.

As in the case of vanishing magnetic charges, one can restrict to
the special case where $M$ is equal to either $|q_L|$ or $|q_R|$.
The two cases are in general inequivalent due to the presence of the
KK monopoles, which give rise to the $Z(\tn)$ orbifold in \cftii,
which we chose, without loss of generality, to act on the
left-moving worldsheet degrees of freedom. The case $M=|q_R|$
corresponds to states that preserve half of the supersymmetry of the
$NS5$ -- KK system. For $M=|q_L|$ one finds extremal,
non-supersymmetric black holes.

In both cases, dimensional reduction of the geometry \cftii\ to
$3+1$ dimensions gives $AdS_2\times S^2$, with
\refs{\GiveonMI,\KutasovZH}
\eqn\rrrrii{R_{\rm AdS}=R_{\rm sphere}={l_s\over2}\sqrt{k}~.}
The dilaton takes the constant value
\eqn\gfour{g_4^2=\sqrt{\tn\tw\over |nw|}~.}
The radii of $S^1$ and $\ts^1$ \minss\ are fixed to the values
\eqn\rsrsii{{R^2\over\alpha'}=\left|{n\over w}\right|~, \qquad
{\tr^2\over\alpha'}={\tw\over\tn}~.}
There are also electric gauge fields on $AdS_2$ and magnetic gauge
fields on $S^2$ associated with $S^1$ and $\ts^1$ respectively. The
entropy \sbh\ reduces in this case to
\eqn\ssii{S_{\rm extremal}=2\pi\sqrt{|nw|\tn\tw}~.}
In the BPS case, this agrees with the entropy of four charge black
holes in $3+1$ dimensions (see \eg\ \BalasubramanianEE).

Further specifying to the case \mqlqr\ leads to the background
\eqn\adsthreeii{(AdS_3)_k\times {SU(2)_{k}\over Z(\tn)_L}\times
\CC_4~,}
which was studied in \KutasovZH.

Note that unlike the pure fundamental string case, here all the
symmetries of the branes are realized in the near-horizon
description. In particular, the $SO(3)$ symmetry corresponding to
rotations in $\IR^3$ in \minss\ is realized in $SU(2)_{k}\over
Z(\tn)_L$, and all unbroken supercharges act non-trivially on the
background. This is natural, since for large magnetic charges the
black holes are large and one can think of \cftii, \adsthreeii\ as a
supergravity background, while the corresponding solution for
strings \cbh, \nearext, is necessarily strongly curved.\foot{ A
related fact is that the asymptotically linear dilaton throat
associated with the strings, \nearhorr, \cbh, is {\it not} obtained
from the one with non-zero magnetic charges by setting the magnetic
charges to zero. Indeed, the background \cftii\ -- \cftnew\ only
makes sense for $\tn\tw\ge 2$.}

\newsec{Heterotic strings}

The worldsheet construction of the heterotic string combines the
right-movers of the superstring with the left-movers of the bosonic
string. Thus it is useful, as a warm-up exercise, to first study the
latter. As usual, due to the infrared instability of the twenty six
dimensional flat spacetime background, reflected in the presence of
the bosonic string tachyon, this case is not entirely physical, but
it is useful as preparation to the heterotic string.

\subsec{Thermodynamic description of highly excited bosonic strings}

The entropy of highly excited perturbative bosonic string states is
linear in the energy, as for type II \bbhh. The inverse Hagedorn
temperature is given by
\eqn\bosbh{\beta_H=4\pi l_s~.}
Following the logic of section 2 one might hope that for a
particular value of $d$ the near-horizon geometry of highly excited
strings in
\eqn\bosback{\IR^{d-1,1}\times \CC_{26-d}}
is given by (compare to \nearhorr)
\eqn\bosnearhor{{SL(2,\IR)_{k_b}\over U(1)}\times \CC_{26-d}~.}
Criticality of the background \bosnearhor\ implies that
\eqn\critbos{{3k_b\over k_b-2}-1=d~.}
One can determine $k_b$ and $d$  by requiring that the thermodynamic
entropy of the two dimensional black hole \bosnearhor\ agree with
the microscopic entropy \bbhh, \bosbh.

The entropy of the bosonic two dimensional black hole \bosnearhor\
is again given by \betah, \hagent\ (with $k\to k_b$). Comparing to
\bosbh\ we conclude that the level $k_b$ must be given by
\eqn\boskk{k_b=4~.}
Note that while this value looks different from the one we found in
the type II case $(k=2)$, it is in fact the same. Before modding out
by $U(1)$, we had in the type II case an $SL(2,\IR)$ WZW model with
$k_b=4$ and three free fermions which contributed $-2$ to the level,
for a total of $k\equiv k_b-2=2$. Here, we have just the bosonic WZW
model, whose level is the same as for type II. This fact will play
an important role in the generalization to the heterotic string.

Plugging \boskk\ into \critbos\ we find that in the bosonic string
our construction works for $d=5$, in contrast to the type II case
where it worked for $d=4$ \geomthreeone. This seems like a problem
for the heterotic case which combines the two, but as we will see
next, there is a natural conjecture that can be made there as well.

\subsec{Heterotic strings as black holes}

The heterotic string compactified to $4+1$ dimensions on a manifold
$\CC_5$ is described by the background
\eqn\fivedhet{\IR^{4,1}\times\CC_5~.}
We would like to find the near-horizon geometry of highly excited
perturbative string states in this background. Following
the discussion of section 2 and the previous subsection, it is
natural to propose that the asymptotic form of this geometry is
(compare to \nearhor)
\eqn\hetnearhor{\IR_t\times\IR_\phi\times\CC_5~.}
The central charge accounting works as follows. For the left-moving
(bosonic) sector, the central charge of $\IR_t\times\IR_\phi$ is
given by
\eqn\leftcc{c=c_t+c_\phi=1+1+3Q^2=5~,}
where we took the slope of the linear dilaton to be $Q=1$, as
before. Thus, the left-moving central charge of \hetnearhor\ is
critical.

For the right-moving (fermionic) sector, in addition to the bosonic
fields $(t,\phi)$ we have their two worldsheet superpartners which
together with \leftcc\ bring the central charge to six. However, the
total central charge in this sector has to be that of $\IR^{4,1}$,
which is $15/2$. Thus, we are missing $3/2$ units of $c_R$.

This is precisely the central charge of three free fermions
$\bar\psi_i$, $i=1,2,3$, which realize a level two right-moving
$SU(2)_R$ current algebra, and an $N=1$ superconformal algebra
needed for consistency of the fermionic string. We add these
right-moving free fermions to \hetnearhor\ and interpret the
resulting $SU(2)$ symmetry as an $SU(2)_R$ subgroup of the
$SO(4)=SU(2)_L\times SU(2)_R$ rotation group of $\IR^4$ \fivedhet.
The resulting background is a natural candidate for the asymptotic
form of the near-horizon geometry of perturbative heterotic strings
in $4+1$ non-compact dimensions.

The heterotic background described above is qualitatively different
from its type II counterpart. It is $4+1$ rather than $3+1$
dimensional, and while in the type II case the $SO(3)$ rotation
group acts trivially on the near-horizon geometry, in the heterotic
string an $SU(2)_R$ subgroup of $SO(4)$ acts non-trivially.

So far we only described the asymptotic form of the near-horizon
geometry. The full background is
\eqn\hetcoset{{SL(2,\IR)_2\over
U(1)}\times\{\bar\psi_1,\bar\psi_2,\bar\psi_3\}\times\CC_5~.}
The first factor in \hetcoset\ is a heterotic coset CFT with
$(0,2)$ superconformal symmetry, which we describe next. Models of
this sort are not well studied, and certainly deserve more
attention. We will consider the more general case where the
left-moving $SL(2,\IR)$ has level $k_b$, and the right-moving one
is a super affine Lie algebra of total level
\eqn\kkb{k=k_b-2~.}
The case \hetcoset\ corresponds to $k=2$, $k_b=4$.

One can define the model algebraically, by specifying the current
that is being gauged. As in the type II and bosonic discussions
above, the left and right moving components of this current are
Cartan subalgebra generators of the left and right-moving
$SL(2,\IR)$, $J_3$ and $\bar J_3$.  However, in the heterotic case
the levels of the left and right-moving $SL(2,\IR)$'s are different
(they are given by $k_b$ and $k$, respectively). Hence, the anomaly
free current is
\eqn\anomfree{\left(J_3,\sqrt{k_b\over k}\bar J_3\right)~.}
The worldsheet superpartner of $\bar J_3$ is gauged as well.

The Euclidean heterotic coset can be described using an analog
of the duality of the cigar to Sine-Liouville in the bosonic case
\refs{\FZZ,\KazakovPM} and to $N=2$ Liouville in the fermionic
case \GiveonPX. We start with the Euclidean version of \hetnearhor,
$\IR_y\times\IR_\phi$ (with $y=it$). The Liouville deformation is a
combination of Sine-Liouville for the left-movers, and $N=2$ Liouville
for the right-movers:
\eqn\sinntwo{\delta\CL=\lambda \overline G_{-\half}e^{-{1\over
Q}(\phi +i\sqrt{k_b\over k}y_L+iy_R)}+{\rm c.c.}~,}
where $Q$ and $k$ are related by \qqqq\ and $\overline G_{-\half}$
is the right-moving supersymmetry generator. As in the other cases
mentioned above \refs{\FZZ\KazakovPM-\GiveonPX}, we expect the
heterotic coset \anomfree\ to be related to the Liouville model
\sinntwo\ by strong-weak coupling duality. It would be interesting
to explore this duality further.

The entropy of the heterotic coset described above is given by a
combination of the fermionic and bosonic ones,
\eqn\enthet{S=\pi l_sM\left(\sqrt{k_b}+\sqrt{k}\right)=\pi
l_sM\left(\sqrt{k+2}+\sqrt{k}\right)~.}
For the case of interest, \hetcoset, $k=2$ and the entropy \enthet\
agrees with that of free heterotic strings. Thus, \hetcoset\ is a
natural candidate for the near-horizon geometry of highly excited
heterotic strings in $4+1$ dimensions \fivedhet.

\subsec{Charged heterotic strings as black holes}

If the geometry \fivedhet\ has the form
\eqn\fivedhets{\IR^{4,1}\times S^1\times\CC_4~,}
one can study the thermodynamics of states with left and
right-moving momentum $(q_L,q_R)$ on $S^1$ as in subsection 2.2.
For large oscillator levels $N_L$ and/or  $N_R$,
the entropy of free heterotic
strings with mass $M$ and charges $(q_L,q_R)$ is given by
\eqn\smqlqrh{
S=2\pi\sqrt{2}\left(\sqrt{2N_L}+\sqrt{N_R}\right) =\pi l_s\sqrt
2\left(\sqrt{2(M^2-q_L^2)}+\sqrt{M^2-q_R^2}\right)~.}
Note the relative factor of $\sqrt 2$ between the left-moving
(bosonic) and right-moving (fermionic) sectors of the theory
(compare to \smqlqr).

For $q_L=q_R=0$ we proposed in the previous subsection that the black
hole background \hetcoset\ provides a thermodynamic description of
these states. Black holes with generic $(q_L,q_R)$ can be obtained from
the uncharged one as before, and correspond to the background
\eqn\cbhh{{SL(2,\IR)_2\times U(1)\over U(1)}
\times\{\bar\psi_1,\bar\psi_2,\bar\psi_3\}\times\CC_4~.}
The charge to mass ratio of the black hole determines the way the
$U(1)$ in the denominator acts on $SL(2,\IR)_2\times U(1)$ as in
subsection 2.2.

For general $k$, the entropy of the heterotic coset in \cbhh\ is given
by \GiveonMI:
\eqn\entropy{ S=\pi
l_s\left(\sqrt{(k+2)(M^2-q_L^2)}+\sqrt{k(M^2-q_R^2)}\right)~.}
For the special case $k=2$ it  agrees with that of free heterotic
strings \smqlqrh.

In the supersymmetric extremal case \mmqq\ and the
non-supersymmetric one $M=|q_L|$, one can again formally think
about the coset as a sigma model on $AdS_2\times S^1$, whose
properties are given in \rrads\ -- \diltwo\ with $k=2$. A further restriction
to the case \mqlqr\ leads to the heterotic string on
$SL(2,\IR)_2\times\{\bar\psi_1,\bar\psi_2,\bar\psi_3\}\times\CC_4$,
with the three dimensional string coupling given by \gthree.

\subsec{States carrying electric and magnetic charges}

To add magnetic charges we consider the heterotic string in the
$3+1$ dimensional background \minss. In general, the two circles
$S^1$ and $\ts^1$ combine with the left-moving worldsheet fields
into an even, self-dual Narain torus $\Gamma^{2,18}$ (see \eg\
\PolchinskiRQ). We will restrict to the case where this torus
factorizes,
$\Gamma^{2,18}=\Gamma^{1,1}\times\Gamma^{1,1}\times\Gamma^{16}$, and
the spacetime fields associated with $\Gamma^{16}$ are not excited.
As in subsection 2.3, we will study configurations that contain
$\tw$ $NS5$-branes and $\tn$ KK monopoles wrapped around
$S^1\times\CC_4$ and charged magnetically under the gauge fields
associated with the Neveu-Schwarz B--field and metric on $\ts^1$,
respectively.

We are interested in excitations with mass $M\gg M_s$ and momentum
and winding $(n,w)$ around the $S^1$. The analogous system in type
II string theory has the near-horizon geometry \cftii. This
background, properly interpreted (see \KutasovZH\ and the previous
subsections), is the near-horizon geometry in the heterotic case
too. The level $k$ is given in this case by
\eqn\kkk{k=\tn\tw+2~.}
For the right-movers this means that the bosonic $SL(2,\IR)$ and
$SU(2)$ sigma models have current algebras of level $k+2$ and $k-2$
respectively, as well as worldsheet fermions in the adjoint that
complete both levels to $k$. In the left-moving sector, the fermions
are absent, while the bosonic sigma model has the same properties as
for the other worldsheet chirality. The $Z(\tn)$ orbifold acts on
the left-moving, bosonic, side. The entropy corresponding to \cftii\
is in this case given by \entropy, with the value of $k$ given in
terms of the magnetic charges in \kkk.

As in the type II case, there are a number of special cases in which
the background \cftii\ simplifies.
For uncharged excitations
($q_L=q_R=0$) one finds again the near-extremal fivebrane/KK
monopole background \cftnew, with the left/right asymmetry discussed
in the general case.
The entropy \entropy\ takes the Hagedorn form \enthet,
with $k=\tn\tw+2$.

One can also consider the extremal cases $M=|q_R|$, which is BPS,
and $M=|q_L|$, which is not. In both cases, the near-horizon
geometry \cftii\ reduces to $AdS_2\times S^2$  \refs{\KutasovZH,\GiveonZZ,\GiveonMI}.
The sizes of the anti
de-Sitter space and the sphere are again given by \rrrrii, with the
appropriate value of $k$, \kkk. The four dimensional string coupling
takes a form similar to \gfour, with $\tn\tw\to\tn\tw+2$ (due to the
change in level from \kkkii\ to \kkk). The radii of $S^1$ and $\ts^1$
are independent of $k$ and are again given by \rsrsii.

As in the type II case, the geometric features above are obtained by
continuing from a regime in which the gravity approximation is
reliable. For small $(n,w,\tn,\tw)$ one should instead study the
exact CFT \cftii, however, the gravity analysis is still useful for
many purposes. Indeed, the values of the fields found above from the
CFT \cftii\ agree with those obtained in higher derivative gravity
\refs{\BehrndtEQ\LopesCardosoWT\LopesCardosoCV\LopesCardosoUR
\LopesCardosoXN\MohauptMJ\LopesCardosoQM-\LopesCardosoFP,\SenIZ,\MohauptJD}.

The entropy \entropy\ reduces in the extremal cases to:
\eqn\sns{\eqalign{ M=&|q_R|:\qquad S_{\rm
susy}=2\pi\sqrt{|nw|(\tn\tw+4)}~,
\cr M=&|q_L|:\qquad S_{\rm
non-susy}=2\pi\sqrt{|nw|(\tn\tw+2)}~.\cr}}
This should be compared to equation \ssii\ which holds for both
supersymmetric and non-supersymmetric extremal black holes in the
type II case. For BPS black holes, \sns\ agrees with previous
analyses \refs{\BehrndtEQ\LopesCardosoWT\LopesCardosoCV
\LopesCardosoUR\LopesCardosoXN\MohauptMJ
\LopesCardosoQM-\LopesCardosoFP,\SenIZ,\MohauptJD}.
For the non-BPS extremal black holes, it agrees with
\refs{\KrausVZ,\KrausZM,\SahooPM}.

We see that for generic $\tn$, $\tw$ the heterotic construction and
results are very similar to the type II ones. There is however an
important difference. In the heterotic case, setting $\tn\tw=0$ in
\kkk, the level $k$ takes the value $k=2$, for which the heterotic
version of \cftii\ make sense, and it is natural to associate it
with the near-horizon background of heterotic fundamental strings,
as was done in the previous subsections.\foot{According to
\KutasovZH\ the heterotic string in the background of $\tw$
NS5-branes is described by setting $\tn=1$ (and not to zero, as one
may naively expect). Setting $\tn=0$ seems, instead, to correspond
to no fivebranes.} Moreover, in this case the extremal entropy \sns,
or more generally the non-extremal one \entropy, agrees precisely
with that of perturbative heterotic strings with the same quantum
numbers, \smqlqrh, as discussed above.

Finally, note that when $\tn=0$ the radius  of $\ts^1$ at the
horizon \rsrsii\ goes to infinity, in agreement with the proposal
that the corresponding small black holes describe fundamental
heterotic string states in 4+1 dimensions.

\newsec{Discussion}

In this note we proposed a thermodynamic description of perturbative
string states with $M\gg M_s$ in $3+1$ (4+1) dimensional type II
(heterotic and bosonic) string theory in
terms of small black holes in an asymptotically linear dilaton
spacetime. We showed that these black holes have the correct entropy
for both uncharged and electrically charged states and discussed the
generalization to non-zero magnetic charges.

Perturbative strings are believed to be well described by weakly
coupled string theory in flat spacetime. A natural question is what
is the relation of the black hole description proposed here to the
standard approach. A possible answer is the following. String
perturbation theory is known to break down at high energies. A
dramatic manifestation of this is the formation of large black holes
at energies above $M_s/g_s^2$, but weak coupling techniques are
known to break down in scattering at energies large compared to the
string scale as well \refs{\GrossAR,\MendeWT}. Therefore, it is
natural to expect that many high energy properties of fundamental
strings are hard to compute using perturbative string theory, and
are instead captured by the linear dilaton throat geometry.

An example is scattering off a highly excited string. Probes with
vanishing angular momentum can explore the throat associated with
the fundamental strings and the scattering amplitude might receive a
contribution from this region. In order to reproduce the black hole
result from the perturbative S-matrix one has to sum over
contributions in which the massive target is a multi-string state
consisting of an arbitrary number of strings. This inclusive process
might be hard to study using the standard perturbative approach, but
if this is feasible, it would be interesting to compare the results
to those obtained in the two dimensional black hole background. This
may lead to new insights on the black hole information paradox, the
resolution of the black hole singularity and other related issues.

Our discussion also sheds light on the string/black hole
correspondence studied in
\refs{\VenezianoZF\SusskindWS\HorowitzNW-\DamourAW,\KutasovRR,\GiveonJV,\CornalbaHC}.
{}For states with $M\gg M_s/g_s^2$, the gravity approximation is
valid and the appropriate thermodynamic description is in terms of
large black holes with low Hawking temperature in asymptotically
flat spacetime. As the mass decreases, the Hawking temperature
increases. In the correspondence region $M\sim M_s/g_s^2$ the
temperature reaches the string scale and string corrections to the
geometry become significant. Thus, one would expect significant
corrections to the thermodynamics of such string size black holes.

Similarly, the thermodynamics of free strings, which is expected to be valid
for $M\ll M_s/g_s^2$, is expected to receive large corrections as one
approaches the correspondence region, due to gravitational self-interactions
of the strings. Nevertheless, the authors of \refs{\VenezianoZF\SusskindWS\HorowitzNW-\DamourAW}
pointed out that extrapolating the large black hole and free string results to this
regime leads to qualitative agreement. This is known as the string/black hole
correspondence principle.

In Euclidean black hole solutions, the time coordinate approaches
asymptotically a circle of circumference $\beta$. Winding around
this circle is not conserved as strings can unwind near the
Euclidean horizon. In \refs{\KutasovRR,\GiveonJV} it was proposed
that  the sigma model corresponding to such black holes has a
non-zero condensate of the closed string tachyon winding around the
circle. For large black holes this is a small non-perturbative
effect in the black hole sigma model, which influences the physics
in a region of size $l_s$ around the Euclidean horizon. However, as
the Hawking temperature increases, the effects of this tachyon
become more important and eventually, when the Hawking temperature
approaches the Hagedorn temperature, the tachyon becomes massless
and its fluctuations extend all the way to infinity.

The Minkowski analog of the tachyon condensate is a gas of
fundamental strings at the appropriate temperature. The radial
extent of the condensate is a measure of the effective size of such
strings. For large mass it is small due to the gravitational effects
discussed in \refs{\VenezianoZF\SusskindWS\HorowitzNW-\DamourAW}. As
the mass decreases, this size grows since gravity becomes weaker.

For masses in the correspondence region $M\sim M_s/g_s^2$ the strings are strongly
interacting and hence we expect the temperature to be well below the Hagedorn
temperature. Thus, the tachyon is massive and the effective size of generic string
states is of order one, as discussed in \refs{\VenezianoZF\SusskindWS\HorowitzNW-\DamourAW}.
This region is hard to analyze from both the black hole and fundamental string perspectives.

As the mass continues to decrease below the correspondence region,
the temperature continues to grow and eventually, as $g_s^2M/M_s\to
0$, it approaches its limiting value -- the Hagedorn temperature. In
the process, the tachyon in the Euclidean black hole solution
becomes lighter, and its condensate extends farther and farther in
the radial direction. This leads to a smooth crossover between the
black hole and free string behaviors.

For $g_s^2M/M_s\ll1$ a linear dilaton throat develops in a string
size region around the horizon. As $M$ decreases, this throat
becomes larger, and as long as $M\gg M_s$ the string coupling
outside the horizon remains small. In the limit $g_s\to 0$ with
$M/M_s$ large but fixed the temperature approaches the Hagedorn
temperature, while the part of the geometry corresponding to finite
$r$ approaches $\IR^3\times S^1$ (or $\IR^4\times S^1$ in the
heterotic case) and decouples from the linear dilaton throat. One
can think of the two parts of the geometry as providing the
microscopic and thermodynamic descriptions of the relevant states,
respectively. In particular, the entropy of perturbative strings is
given by the BH entropy of the corresponding two dimensional black
hole.

The Euclidean two dimensional black hole, which we proposed as a
description of the near-horizon geometry of small black holes, is
known to contain a condensate of a tachyon winding around Euclidean
time \refs{\FZZ\KazakovPM-\GiveonPX}. For finite value of the
asymptotic string coupling, this geometry should attach smoothly to
the flat spacetime at infinity. Thus, it is natural to expect that,
at least for small black holes, the expectation value of the winding
tachyon is non-zero at large distances from the horizon as well, in
agreement with the proposal of \refs{\KutasovRR,\GiveonJV}.
Decreasing the BH temperature (or increasing the mass towards the
large black hole regime) should not change the fact that the tachyon
condensate is non-trivial.

\bigskip
\noindent{\bf Acknowledgements:} We thank O. Aharony,
A.~Dabholkar, O. Lunin, E. Martinec, M. Porrati and G. Veneziano
for discussions. This work is supported in part by the BSF --
American-Israel Bi-National Science Foundation. AG is supported in
part by the Israel Science Foundation, EU grant
MRTN-CT-2004-512194, DIP grant H.52, and the Einstein Center at
the Hebrew University. The work of DK is supported in part by DOE
grant DE-FG02-90ER40560 and the National Science Foundation under
Grant 0529954. AG thanks the EFI at the University of Chicago for
hospitality. DK thanks the Aspen Center for Physics, Weizmann
Institute and Hebrew University for hospitality during the course
of this work.

\listrefs
\end